\documentclass{elsart}
\usepackage{graphics}
\usepackage{amssymb}
\usepackage{rotating}

\begin{document}

\begin{frontmatter}
\title{Fixed boundary conditions analysis of the 3d Gonihedric Ising
       model with $\kappa=0$}
\author[UAB]{M.~Baig,}
\author[UAB]{J.~Clua,}
\author[WATT]{D.A.~Johnston}
\author[FABRA1,FABRA2]{and R.~Villanova}

\address[UAB]{IFAE, Universitat Aut{\`o}noma de Barcelona,
             08193 Bellaterra (Barcelona) Spain}
\address[WATT]{Dept. of Mathematics, Heriot-Watt University,
             Edinburgh, EH14 4AS,
             Scotland}
\address[FABRA1]{Matem{\`a}tiques Aplicades, DEE, Universitat Pompeu Fabra,
             Barcelona, Spain}
\address[FABRA2]{{\bf Corresponding author:} \\
                 Ramon Villanova, Matem{\`a}tiques Aplicades, DEE, \\
                 Universitat Pompeu Fabra, C/ Trias Fargas, 25-27 \\
                 08005 Barcelona, Spain  \\
                 phone: 34 93 542 2733, $\; \,$ FAX: 34 93 542 2296\\
                 E-mail: ramon.villanova@upf.edu }

\begin{abstract}
 The Gonihedric Ising model is a particular case of the class of
 models defined by Savvidy and Wegner intended as discrete versions of
 string theories on cubic lattices. In this paper we perform a high
 statistics analysis of the phase transition exhibited by the 3d
 Gonihedric Ising model with $k=0$ in the light of a set of recently stated
 scaling laws applicable to first order phase transitions with fixed
 boundary conditions.
 Even though qualitative evidence was
 presented in a previous paper to support
 the existence of a first order phase
 transition at $k=0$, only now are we capable of pinpointing
 the transition inverse temperature at $\beta_c = 0.54757(63)$
 and of checking the scaling of standard observables.

\end{abstract}


\end{frontmatter}

\section{Introduction}

In a recent paper~\cite{BaigV} we have studied the effects of
freezing the boundaries in a Monte Carlo simulation near a first
order phase transition. More specifically, we checked (and
postulated one of) the scaling laws governing the critical regime
of the transition by means of a Monte Carlo simulation of the 2d,
8-state spin Potts model. These new scaling laws, theoretically
analyzed by C. Borgs and R. Koteck\'{y} and by I. Medved~\cite{Borgs},
imply a major change in the critical behavior analysis.

The MC simulation of a system with fixed boundary conditions
(F.B.C.) instead of the standard periodic ones (P.B.C.) is more
than a simple academic exercise. Indeed, the numerical analysis of
the 3d Gonihedric Ising model requires fixing the spins of some
internal planes. If periodic boundary conditions are adopted, the fixing of
these internal planes is just equivalent to the simulation of the
system in a box with fixed boundary conditions. For this reason, the
Gonihedric Ising model with $\kappa=0$, which manifests a first order
phase transition~\cite{BaigEJM}, needs to be reanalyzed in the
light of the appropriate scaling laws. Moreover, in our recent
paper~\cite{BaigV}, the new scaling laws were checked for a two
dimensional system, so the 3d Gonihedric Ising model offers the opportunity to
extend their verification to 3d lattices.

In the present paper we perform a high statistics study of the 3d
Gonihedric Ising model with $\kappa=0$ at the transition
point on lattices up
to $20^3$. Our analysis of the scaling behavior of some standard
thermodynamical magnitudes (specific heat, susceptibility and
energetic Binder cumulant) confirms the above-mentioned scaling
laws and shows the importance of applying the correct scaling
forms when fixed boundary conditions are present.

This letter is divided as follows. A brief summary of the
Gonihedric Ising model is contained in Sec.~2. The scaling laws
for first order phase transitions are stated in Sec.~3, comparing
the laws for fixed boundary conditions with their periodic
counterparts. Sec.~4 and Sec.~5 are devoted to the numerical
simulation and analysis of results and Sec.~6
summarizes the conclusions of our work.

\section{The Gonihedric Ising model at $\kappa=0$}
Adding extended range interactions, particularly with different
sign couplings, to the standard Ising model in two and three
dimensions gives a very rich~\cite{Cappi} phase structure.  One
particular class of models with such extended interactions, the
so-called Gonihedric Ising models, have recently aroused interest
because of their putative connection with random surface models
and strings.  The original discretized random surface model was
developed by Savvidy et al.~\cite{Savvidy} with the action
\begin{equation}
S = {1 \over 2} \sum_{<ij>} | \vec X_i - \vec X_j | \theta
(\alpha_{ij}), \label{savvidy}
\end{equation}
where the sum is over the edges of some triangulated surface,
$\theta(\alpha_{ij}) = | \pi - \alpha_{ij} |^{\zeta}$, $\zeta$ is
some exponent, and $\alpha_{ij}$ is the dihedral angle between
neighbouring triangles with common link $<ij>$. It was christened the
Gonihedric string model.

The above action was translated to {\em plaquette} surfaces by
Savvidy and Wegner~\cite{Wegner,Pietig} who rewrote the resulting
theory as a generalized Ising model by using the geometrical spin
cluster boundaries to define the plaquette surfaces. In view of
its relation to the Gonihedric string model, this new action was
named the Gonihedric Ising model. In what follows we shall consider
the three dimensional version of this model, whose Hamiltonian
contains nearest neighbour ($<i,j>$), next to nearest neighbour
($<<i,j>>$) and round a plaquette ($[i,j,k,l]$) terms
\begin{equation}
H= 2 \kappa \sum_{<ij>}^{ }\sigma_{i} \sigma_{j} -
\frac{\kappa}{2}\sum_{<<i,j>>}^{ }\sigma_{i} \sigma_{j}+
\frac{1-\kappa}{2}\sum_{[i,j,k,l]}^{ }\sigma_{i}
\sigma_{j}\sigma_{k} \sigma_{l}. \label{Gonihedric_action}
\end{equation}
For generic couplings the spin clusters in the above Hamiltonian
generate a gas of surfaces with energy contributions from area,
extrinsic curvature and self-intersections~\cite{Karowski}. A
noteworthy feature of the particular ratio of couplings in
Eq.~(\ref{Gonihedric_action}) is the flip symmetry which is not
present in the generic case. It is possible to flip any plane of
spins at zero energy cost when $T=0$, so the zero temperature
ground state is degenerate, with any layered configuration being
equivalent to the ferromagnetic state. A low temperature expansion
shows that this symmetry is lost when $T \ne 0$ and $\kappa \ne 0$
~\cite{Pietig}. $\kappa=0$ however constitutes a special case --
the flip symmetry remains even at finite temperature.

There is agreement on the  phase structure of the Hamiltonian in
Eq.~(\ref{Gonihedric_action}) from both Monte Carlo
 simulations and cluster-variational
(CVPAM) methods: when $\kappa > 0$ there is a single continuous
transition from a paramagnetic high temperature phase to (with
appropriate boundary conditions in the Monte Carlo case) a
ferromagnetic phase. The simulations of Ref.~\cite{Johnston} used
fixed boundary conditions in order to define a magnetic order
parameter; the reason was that it was found that with the use of
standard periodic boundary conditions flipped spin layers, with
arbitrary interlayer spacings, made it unfeasible.

The nature of the transition for $\kappa \sim 0$ was then
investigated in Ref.~\cite{BaigEJM}. A zero temperature
analysis~\cite{Johnston} shows that there is a further
``antiferromagnetic'' symmetry in the ground state when
$\kappa=0$, which is already apparent from the Hamiltonian itself.
This extra symmetry, and the persistence of flip symmetries at
non-zero $T$ suggest that $\kappa=0$ is a special point in the
space of Hamiltonians Eq.(\ref{Gonihedric_action}). Even though
the results of Ref.~\cite{BaigEJM} suggested the presence of a
first order phase transition at $\kappa=0$, a complete finite size
analysis of the transition was not performed at that time for want
of a better knowledge of the scaling laws applicable with fixed
boundary conditions.

\section{The new scaling laws for frozen boundaries}

As mentioned in the introduction, the scaling laws applicable to
systems simulated with fixed boundary conditions were deduced and
studied in Ref.~\cite{BaigV,Borgs}. The numerical analysis of
Ref.~\cite{BaigV} was performed on the 2d 8-state Potts model.
Since the difference between the corresponding scaling laws for
fixed and periodic boundary conditions are highly volume-dependent, in
addition to its intrinsic interest the simulation of the 3d
Gonihedric Ising model is a good testing ground for the new scaling laws
on a 3d lattice.

A main feature of the F.B.C. simulations is the shift of the
infinite volume inverse temperature by a $1/L$ correction term,
caused by surface effects, instead of the $1/L^d$ correction term
due to volume effects seen in the periodic case.
The same change in the shift is also observed
for the energetic Binder parameter with fixed boundary conditions.

Moreover, the surface corrections to the volume scaling of the
specific heat and the susceptibility become of order $L^{d-1}$
in the fixed case
instead of the almost negligible $1/L^d$.

Table~\ref{tab:scal-laws} summarizes the scaling laws for a first
order phase transition for both periodic and fixed boundary
conditions.

\section{Numerical simulation}
As we have already noted, the flip symmetry poses something of a
problem when carrying out simulations since it means that a simple
ferromagnetic order parameter
\begin{equation}
m = \left< {1 \over L^3} \sum_i \sigma_i \right>. \label{order}
\end{equation}
will be zero, because of the observed layered nature of the
ordered state. Staggered magnetizations are of no use since the
inter layer spacing can be arbitrary. On a finite lattice it is
possible, however, to force the model into the ferromagnetic
ground state by fixing sufficient perpendicular spin planes,
either internally if P.B.C. are used or on the boundaries of the
lattice: both possibilities being exactly equivalent.

As in our previous work~\cite{BaigEJM}, we choose to fix internal
planes of spins in the lattice, while retaining the periodic
boundary conditions. This has the desired effect of picking out
the ferromagnetic ground state. We can therefore still employ the
simple order parameter of Eq.~(\ref{order}). For $\kappa=0$ the
Hamiltonian we simulate is\footnote{It is perhaps worth
emphasizing that spins live on the vertices of the cubic lattice
rather than on the links, so the model of Eq.~(\ref{action}) is
{\em not} the three dimensional $Z_2$ gauge model that is dual to
the three dimensional Ising model.}
\begin{equation}
H= \frac{1}{2}\sum_{[i,j,k,l]}^{ }\sigma_{i} \sigma_{j}\sigma_{k}
\sigma_{l}. \label{action}
\end{equation}
Table~\ref{tab:statistics} summarizes the details of the
simulations that have been performed from $L=10$ up to $L=20$. The
lattice updating used a simple Metropolis algorithm. The number of
production Monte Carlo sweeps varies from $n_{\rm
prod}=20~000~000$ for $L=10$, to $n_{\rm prod}=200~000~000$ for
$L=20$. We took measurements of the energy and the magnetization
only every $n_{\rm flip}=4$ or $n_{\rm flip}=8$ sweeps, and,
consequently, the number of total measurements per run is $n_{\rm
meas}=n_{\rm prod}/n_{\rm flip}$. We left at least $21 \, n_{\rm
flip} \tau_{\rm e}$ thermalization sweeps before taking
measurements~\cite{Sokal}. To estimate the autocorrelation time of
energy measurements $\tau_{\rm e}$, we use the fact that
$\tau_{\rm e}$ enters the error estimate $\epsilon_{\rm
JK}=\sqrt{2\; \tau_{\rm e}/n_{\rm meas}} \;\, \epsilon_{\rm
naive}$ for the mean energy $<E>$ of $n_{\rm meas}$ correlated
energy measurements of variance
\begin{equation}
 \epsilon_{\rm naive}^2=\sum_{j=1}^{n_{\rm meas}}(<E>-E_j)^2/(n_{\rm meas}-1).
\end{equation}
The ``true" error estimate $\epsilon_{\rm JK}$ is obtained
splitting the energy time-series into 50 bins, which were in their
turn jackknived~\cite{jack} to decrease the bias in the analysis.

\begin{figure}[p]
  \vspace*{20cm}
  \includegraphics{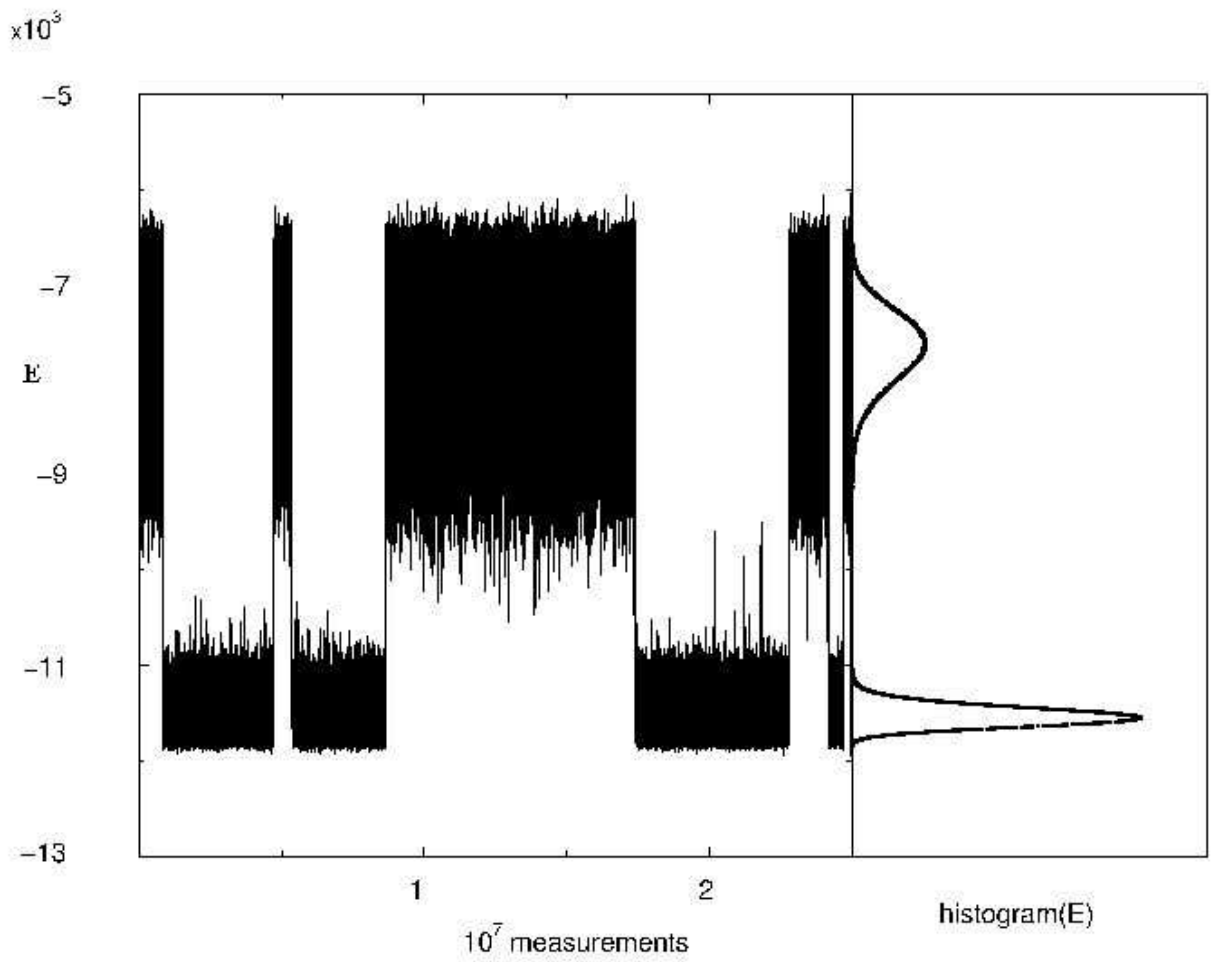}
  \caption{Energy time series and corresponding energy histogram
          for $L=20$ and $\beta_{MC}=0.5064$}
  \label{fig:time-series}
  \end{figure}

In Fig.~\ref{fig:time-series} we present the energy time-series
for the $L=20$ and $\beta_{MC}=0.5064$ simulation run. The
expected characteristic behaviour of a first order phase
transition can be clearly seen. The system remains in one of the
two coexisting phases for a long period of time. The energy
histogram for the full series is also presented in the Figure. The
similar height of the two peaks confirms that the simulation was
performed very near the pseudo-critical inverse temperature.

In addition to the qualitative analysis of the histograms, we have
computed the specific heat, magnetic susceptibility and the
energetic Binder parameter at nearby values of $\beta_{MC}$ by
means of standard reweighting techniques~\cite{Swendsen}. These
observables are defined as
\begin{eqnarray}
C(\beta) & = & \frac{\beta^{2}}{V}(\langle E^{2}\rangle-\langle
E\rangle^2), \\
\chi(\beta) & = & \frac{\beta^{2}}{V}(\langle M^{2}\rangle-\langle
M\rangle^2), \\
B(\beta) & = & 1-\frac{\langle E^4 \rangle}{3 \langle
E^2\rangle^2}.
\end{eqnarray}
In Table~\ref{tab:extrema} we show the extrema of the magnitudes
defined above, together with their pseudo-critical inverse
temperatures. The error bars of these quantities have been
estimated splitting the time-series data into 50 bins, which were
then
jackknived to decrease the bias in the analysis of reweighted
data.

\section{Analysis of results}

Once we have the results from the numerical simulation on finite
lattices, we can proceed to analyze the data by fitting to the scaling laws
of Table~\ref{tab:scal-laws}.

\begin{figure}[p]
 \vspace*{9cm}
   \includegraphics{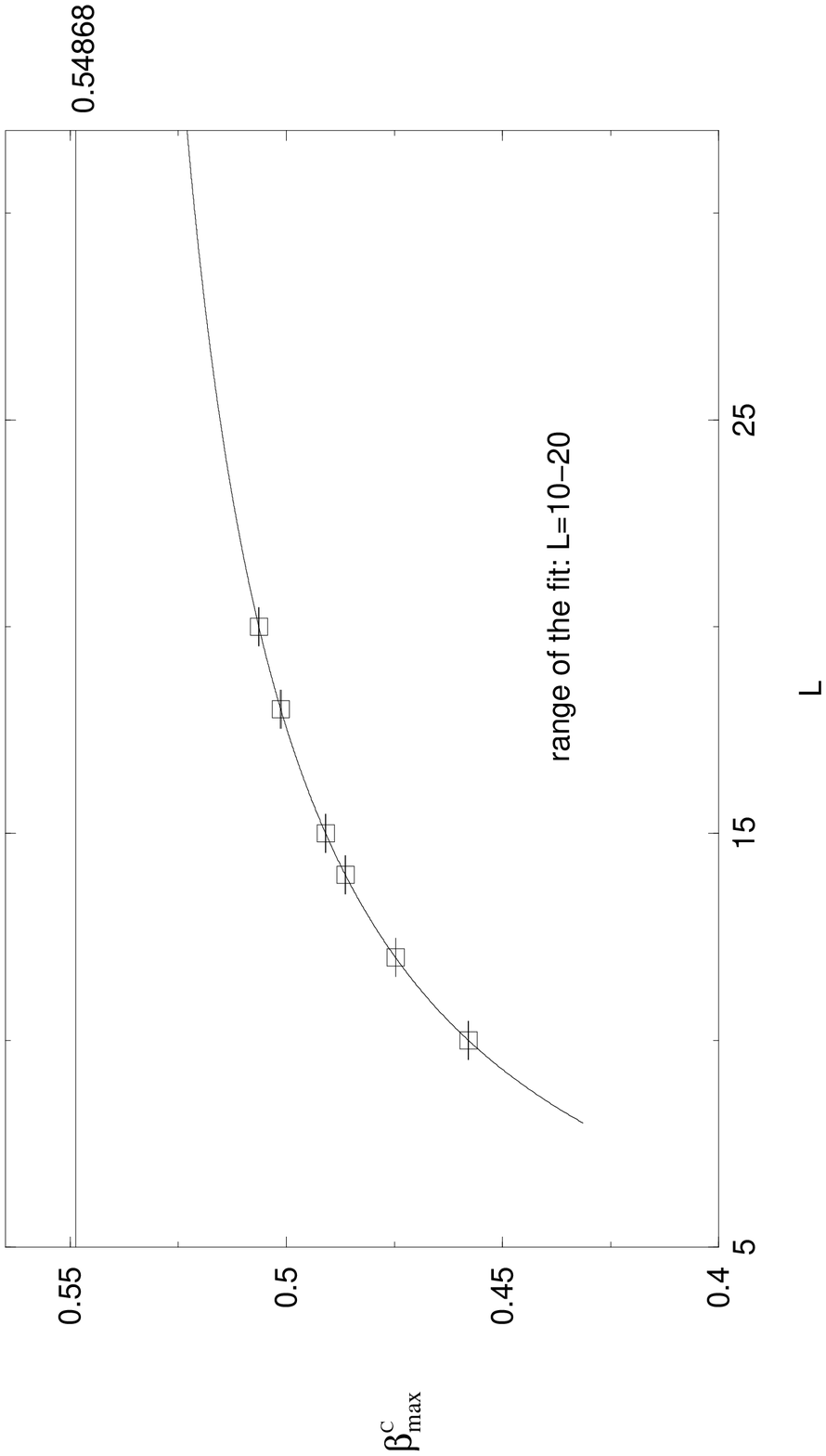}
 \caption{Finite-size scaling analysis of the pseudo-critical
  $\beta^{C}_{\rm max}$ in the range $L=10-20$ by means of the ansatz
  $\beta^{C}_{\rm max}(L)=\beta_c + a_1/L + a_2/L^2$. The
  infinite volume critical point obtained from the fit is
  $\beta_c=0.54868(34)$, with a goodness-of-fit $Q=0.89$.}
  \label{fig:beta-fit}
\end{figure}

In Table~\ref{tab:beta-fits} we show the results of fitting the
pseudo-critical $\beta$s of $C_{\rm max}$, $\chi_{\rm max}$ and
$B_{\rm min}$ to the ansatz
\begin{equation}
  \beta_{\rm max}(L) = \beta_c + \frac{a_1}{L} + \frac{a_2}{L^2}
\end{equation}
suggested by the finite-size scaling laws presented in
Table~\ref{tab:scal-laws}. For $\chi_{\rm max}$ and $B_{\rm min}$
the fits were rather poor if $L=10$ was included, so
it was discarded. For $C_{\rm max}$ both sets $L=10-20$ and
$L=12-20$ were fitted. Focusing on the $L=12-20$ fits, we can
discern only very minor differences in the estimated
$\beta_c$ depending on the
observable used to extract it. These are so small that we can safely
average to obtain
\begin{equation}
    \beta_c = 0.54757 \pm 0.00063
\end{equation}
Since the $\beta_c$'s extracted from the three observables were
not independent, we have kept the error bar common to them all. In
Fig.~\ref{fig:beta-fit} we depict the fit for $\beta^C_{\rm
max}(L)$ in the range $L=10-20$. The error bars in the Figure are
so small that they show up only as horizontal dashes.

The results of the fits to the specific heat and susceptibility
maxima, $C_{\rm max}$ and $\chi_{\rm max}$, together with the
energetic Binder parameter minimum are summarized in
Table~\ref{tab:extrema_fits}. The goodness-of-fit, $Q$, is
excellent for the three observables.

Note that the surface correction coefficients $a_1$ and $b_1$
are, in absolute value, from one to two orders of magnitude larger
than the coefficients $a_2$ and $b_2$ of the dominant
contribution $V=L^3$. It is precisely this fact which makes it necessary to
use the scaling ansatz $C_{\rm max}(L) = a_0 + a_1 \, L^2 +
a_2 \, L^3$, and allows us to estimate the corrections to the
leading term.

\section{Conclusions}
We have performed a numerical simulation of the 3d Gonihedric
Ising model at $\kappa=0$ in order to determine the thermodynamic
characteristics of its phase transition. Previous analysis
suggested the existence of a first order phase transition, but a
complete finite size analysis of the transition was not carried out. The
special features of this model, which requires a simulation where
three perpendicular spin planes need to be fixed during the
simulation, do not allow a direct application of the standard
finite size scaling laws for
periodic boundary conditions at a first order transition. In fact, to
keep these planes fixed is equivalent to performing a simulation with
fixed boundary conditions (F.B.C.), giving rise to the need for a
different set of scaling laws. They were reviewed in Sec.~3.
Our numerical analysis of the thermodynamic quantities has shown
that the critical behavior of the 3d Gonihedric Ising model is perfectly
described in terms of F.B.C. scaling laws. As a result of this
work, we have been able to accurately determine the inverse critical
temperature of the model, i.e. $\beta_c = 0.54757(63)$.
Furthermore, our simulation has extended the verification of the
F.B.C. scaling laws to a three dimensional lattice model.

\vspace{0.8cm}

{\bf Acknowledgements}

M.B. and R.V. acknowledge financial support from MCyT project {\it
BFM 2002-02588} and CIRIT project {\it SGR-00185}, D.J.
acknowledges the partial support of EC network grant {\it
HPRN-CT-1999-00161}.

\newpage

\begin{table}[h]
\centering \caption[{\em Nothing.}] {{\em Scaling laws for
Periodic {\em versus} Fixed Boundary Conditions.}} \vspace{3ex}
\begin{tabular}{|l|c|c|} \hline
    & P.B.C. & F.B.C. \\ \hline
  $\beta_{c}^{peaks}(L) =$ & $\beta_{c}(\infty)+ \frac{\theta_1}{L^{d}}+O(\frac{1}{L^{2d}})$ &
   $\beta_{c}(\infty)+ \frac{a_1}{L}+O(\frac{1}{L^{2}})$ \\
 $C_{max}(L) =$ & $\gamma_{0}+\gamma_{2}L^{d}+O(\frac{1}{L^{d}})$ &
   $c_{0}+c_{2}L^{d}+O(L^{d-1})$ \\
$\chi_{max}(L)= $ &
$\delta_{0}+\delta_{2}L^{d}+O(\frac{1}{L^{d}})$ &
   $e_{0}+e_{2}L^{d}+O(L^{d-1})$ \\
$B_{min}(L) = $ &
$\Phi_{0}+\frac{\Phi_{1}}{L^{d}}+O(\frac{1}{L^{2d}})$ &
$B_{0}+\frac{B_{1}}{L}+O(\frac{1}{L^{2}})$\\ \hline
\end{tabular}
\label{tab:scal-laws}
\end{table}

\vspace{2.5cm}

%
%
\begin{table}[htbp]
\centering \caption[{\em Nothing.}]
 {{\em  Monte Carlo parameters of the simulation. $L^3$ is the
        lattice size, $n_{\rm therm}$ the number of Monte Carlo
        sweeps during thermalization (in thousands), and $n_{\rm prod}$ the number
        of production runs (in millions). Measurements were taken every $n_{\rm flip}=4$
        Monte Carlo sweeps for all the simulations, except the latest;
        the number of bins was 50.}}
\vspace{3ex}
\begin{tabular}{r|lrrrrrrr}
\multicolumn{1}{c|}{$L$}  & \multicolumn{1}{c}{$\beta_{MC}$} &
\multicolumn{1}{c}{$n_{\rm therm}$} & \multicolumn{1}{c}{$n_{\rm
prod}$}   &
 \multicolumn{1}{c}{$n_{\rm flip}$} &
\multicolumn{1}{c}{$\tau_{\bf \rm e}$} &
\multicolumn{1}{c}{$\frac{n_{\rm therm}/n_{\rm flip}}{\tau_{\bf
\rm e}}$} & \multicolumn{1}{c}{$\frac{n_{\rm prod}/n_{\rm flip}}{2
\, \tau_{\bf \rm e}}$} \\
\hline
10 & 0.4580 & 500  &  20  & 4 &    25& 5 000 &  100 000 \\ \\
12 & 0.4748 & 500  &  20  & 4 &    45& 2 778 &   55 556 \\ \\
14 & 0.4864 & 500  &  20  & 4 &   278&   450 &    8 993 \\ \\
15 & 0.4910 & 500  &  20  & 4 & 1 011&   124 &    2 473 \\ \\
18 & 0.5013 & 2 500 &  22  & 4 & 24 871&   25 &      111 \\ \\
20 & 0.5064 & 36 500 & 200 & 8 &216 098 & 21 &       58 \\ \\
 \end{tabular}
\label{tab:statistics}
\end{table}

\newpage

%
%
\begin{table}[htbp]
\centering \caption[{\em Nothing.}]
 {{\em Extrema for the (finite lattice) specific heat, $C_{\rm max}$,
      the susceptibility, $\chi_{\rm max}$, and the energetic Binder
      parameter, $B_{\rm min}$, together with their respective
      pseudo-critical inverse temperatures.}}
\vspace{3ex}
\begin{tabular}{r|lr|lr|ll}
\multicolumn{1}{c|}{$L$}  & \multicolumn{1}{c}{$\beta_{\rm
max}^{C}$} & \multicolumn{1}{c|}{$C_{\rm max}$} &
\multicolumn{1}{c}{$\beta_{\rm max}^{\chi}$} &
\multicolumn{1}{c|}{$\chi_{\rm max}$}    &
\multicolumn{1}{c}{$\beta_{\rm min}^{B}$} &
\multicolumn{1}{c}{$B_{\rm min}$}  \\
\hline
   10  & 0.457919(21) &  5.6945(79) & 0.456842(22) &
         7.042(12)  & 0.455064(21) & 0.638537(47) \\
       &  &  &  &  & & \\
   12  & 0.474753(16) & 12.120(21) & 0.474470(15) &
         16.305(31) & 0.473468(16) & 0.635656(61) \\
       &  &  &  &  & & \\
   14  & 0.486349(21) & 25.172(45) & 0.486275(21) &
         36.264(73) & 0.485647(21) & 0.628430(85) \\
       &  &  &  &  & & \\
   15  & 0.490922(30) & 34.900(76) & 0.490884(30) &
         51.91(13)  & 0.490374(30) & 0.62432(12) \\
       &  &  &  &  & & \\
   18  & 0.501280(72) & 78.99(38) $\:$ & 0.501273(72) &
         128.10(69) & 0.500979(72) & 0.61246(36)  \\
       &  &  &  &  & & \\
   20  & 0.506366(69) & 121.57(52) $\:$ & 0.506364(69) &
         206.61(99) & 0.506149(69) & 0.60620(36)  \\
\end{tabular}
\label{tab:extrema}
\end{table}

\newpage

%
%
%
%

\begin{sidewaystable}
\begin{small}
\centering \caption[{\em Nothing.}]
 {\em  Pseudo-critical inverse temperature fits. Q is the
        goodness-of-fit. }
\vspace{3ex} \hspace*{0cm}\begin{tabular}{r|lrll|lrll|lrll}
\multicolumn{1}{c|}{range L's}  &
\multicolumn{4}{c|}{$\beta^C_{\rm max}(L)=\beta_c+a_1/L+a_2/L^2$}
& \multicolumn{4}{c|}{$\beta^{\chi}_{\rm
max}(L)=\beta_c+d_1/L+d_2/L^2$} & \multicolumn{4}{c}{$\beta^B_{\rm
max}(L)=\beta_c+e_1/L+e_2/L^2$}  \\ \hline & \multicolumn{1}{c}{Q}
& \multicolumn{1}{c}{$\beta_c$} & \multicolumn{1}{c}{$a_1$} &
\multicolumn{1}{c|}{$a_2$} &
\multicolumn{1}{c}{Q}&\multicolumn{1}{c}{$\beta_c$}&\multicolumn{1}{c}{$a_1$}
&\multicolumn{1}{c|}{$a_2$}&
\multicolumn{1}{c}{Q}&\multicolumn{1}{c}{$\beta_c$}&\multicolumn{1}{c}{$a_1$}
&\multicolumn{1}{c}{$a_2$} \\ \hline \hline
              &  &  &  &  &  &  &   &  & & & &  \\
  10 -- 20 & 0.89 & 0.54868(34) & -0.7848(84) & -1.229(51)&  &
   & & & & & &  \\
              &  &  &  &  &  &  &   &  &  \\
  12 -- 20 & 0.73 & 0.54867(63) & -0.785(18) & -1.23(12)& 0.85 &
                    0.54730(63) & -0.736(18) & -1.65(12)& 0.86 &
                    0.54674(63) & -0.711(18) & -2.02(12) \\
            &  &  &  &  &  &  &   &  & & & &  \\
\end{tabular}
\end{small}
\label{tab:beta-fits}
\end{sidewaystable}

%
%

\begin{sidewaystable}
\begin{small}
\centering \caption[{\em Nothing.}]
 {\em  Fits on the extrema of $C_{\rm max}$, $\chi_{\rm max}$ and
       $B_{\rm min}$.}
\vspace{3ex} \hspace*{0cm}\begin{tabular}{r|llll|llll|llll}
\multicolumn{1}{c|}{range L's} & \multicolumn{4}{c|}{$C_{\rm
max}(L)    = a_0 + a_1 \, L^2 + a_2 \, L^3$}  &
\multicolumn{4}{c|}{$\chi_{\rm max}(L) = b_0 + b_1 \, L^2 + b_2 \,
L^3 $} & \multicolumn{4}{c} {$B_{\rm min}(L) =  B_0 + B_1 / L +
B_2 / L^2$} \\ \hline & \multicolumn{1}{c}{Q} &
\multicolumn{1}{c}{$a_0$} & \multicolumn{1}{c}{$a_1$}&
  \multicolumn{1}{c|}{$a_2$} &
 \multicolumn{1}{c}{Q} & \multicolumn{1}{c}{$b_0$} & \multicolumn{1}{c}{$b_1$}&
  \multicolumn{1}{c|}{$b_2$} &
 \multicolumn{1}{c}{Q} & \multicolumn{1}{c}{$B_0$} & \multicolumn{1}{c}{$B_1$}&
  \multicolumn{1}{c}{$B_2$} \\ \hline \hline
 &  &  &  &  &  &  &  & & & & & \\
 10 -- 20  & 0.16 & 14.43(17) & -0.4434(36) & 0.03561(20)&  & & &
              & 0.014 & 0.4992(15) & 2.851(35) & -14.57(21) \\
 &  &  &  &  &  &  &  & & & & & \\
 12 -- 20  & 0.098& 14.79(52) & -0.4491(83)& 0.03587(40)& 0.62& 39.92(92)& -1.035(15)& 0.07257(73)
                        & 0.21 & 0.5065(30)& 2.643(84) & -13.12(57) \\

\end{tabular}
\end{small}
\label{tab:extrema_fits}
\end{sidewaystable}

\newpage

\begin{thebibliography}{99}

\bibitem{BaigV} M.~Baig and R.~Villanova. Phys. Rev. {\bf B 65}, 094428
                (2002).
\bibitem{Borgs} C.~Borgs and R.~Koteck\'{y}, Journ. Stat. Phys. {\bf
                61}, 79 (1990); ibidem {\bf 79}, 43 (1995).
                I. Medved, diploma Thesis. Charles University. Prague.
\bibitem{BaigEJM} M.~Baig, D.~Espriu, D.A.~Johnston and R.P.K.C.~Malmini, J. Phys. A
                {\bf 30}, 405 (1997); ibidem {\bf 30}, 7695 (1997).
\bibitem{Cappi} A.~Cappi, P.~Colangelo, G.~Gonnella, and A.~Maritan, {\em Nucl. Phys.}
                {\bf B370}, 659 (1992). W.~Selke {\em Phys. Rep.} {\bf 170}, 213 (1988).
\bibitem{Savvidy} R.V.~Ambartsumian, G.S.~Sukiasian, G.K.~Savvidi, and
                K.G.~Savvidy, {\em Phys. Lett.} {\bf B275}, 99 (1992). G.K.~Savvidy
                and K.G.~ Savvidy,  {\em Mod. Phys. Lett.} {\bf A8}, 2963 (1993).
                G.K.~Savvidy and K.G.~Savvidy, {\em Int. J. Mod. Phys.} {\bf A8}, 3993
                (1993).
\bibitem{Wegner} G.K.~Savvidy and F.J.~Wegner, {\em  Nucl. Phys.} {\bf B413}, 605
                (1994).  G.K.~Savvidy and K.G.~Savvidy, {\em Phys. Lett.} {\bf B324}, 72
                (1994); ibidem {\bf B337}, 333 (1994). G.K.~Savvidy, K.G.~Savvidy, and P.G.~Savvidy, {\em Phys.Lett.}
                {\bf A221}, 233 (1996). G.K.~Savvidy, K.G.~Savvidy and F.J.~Wegner,{\em Nucl. Phys.}
                {\bf B443}, 565 (1995).
\bibitem{Pietig} G.K.~Bathas, E.~Floratos, G.K.~Savvidy, and K.G.~Savvidy,
                {\em Mod. Phys. Lett.} {\bf A 10}, 2695 (1995). R.~Pietig and
                F.J.~Wegner, {\em Nucl. Phys.} {\bf B466}, 513 (1996).
\bibitem{Karowski} T.~Sterling and J.~Greensite, {\em Phys. Lett.} {\bf B121}, 345 (1983).
                M.~Karowski and H.J.~Thun, {\em Phys. Rev. Lett.} {\bf 54}, 2556 (1985).
                M.~Karowski {\em J. Phys.} {\bf A19} (1986) 3375.
\bibitem{Johnston} D.A.~Johnston and R.P.K.C.~Malmini, {\em Phys. Lett.} {\bf B378}, 87 (1996).
\bibitem{Sokal} A.D.~Sokal in {\em Monte Carlo Methods in Statistical
                Mechanics: Foundations and New Algorithms\/}, lecture
                notes, Cours de Troisi\`eme Cycle de la Physique en Suisse
                Romande, Lausanne (1989). N.~Madras and A.D.~Sokal, J. Stat. Phys. {\bf 50}, 109
                (1988). W.~Janke, {\em Monte Carlo Simulations of Spin Systems\/}, in: {\em Computational
                Physics: Selected Methods -- Simple Exercises --Serious
                Applications\/}, eds. K.H.~Hoffmann and M.~Schreiber
                (Springer, Berlin, 1996), p.~10.
\bibitem{jack}  R.G.~Miller, Biometrika {\bf 61}, 1 (1974). B.~Efron,
                {\em The Jackknife, the Bootstrap and other Resampling Plans\/}
                (SIAM, Philadelphia, PA, 1982).
\bibitem{Swendsen} A.M.~Ferrenberg and R.H.~Swendsen, Phys. Rev. Lett.
                {\bf 61}, 2635 (1988).
\end{thebibliography}
\end{document}